\documentclass[epj,nopacs]{svjour}
\usepackage{graphics}
\usepackage{amsmath}
\usepackage{amssymb}

\begin{document}
\title{Left-right asymmetry for pion and kaon production in the semi-inclusive deep inelastic
scattering process}
\author{Bo Sun \and Jun She \and Bing Zhang \and Ya-Jun Mao \and Bo-Qiang Ma}

\institute{School of Physics and State Key Laboratory of Nuclear
Physics and Technology, Peking University, Beijing 100871, China,
\email{maoyj@pku.edu.cn}\\ \email{mabq@phy.pku.edu.cn}}
\date{Received: date / Revised version: date}
%
\abstract{ We analyze the left-right asymmetry in the semi-inclusive
deep inelastic scattering~(SIDIS) process without introducing any
weighting functions. With the current theoretical understanding, we
find that the Sivers effect plays a key role in our analysis. We use
the latest parametrization of the Sivers and fragmentation functions
to reanalyze the $\pi^\pm$ production process and find that the
results are sensitive to the parametrization. We also extend our
calculation on the $K^{\pm}$ production, which can help us know more
about the Sivers distribution of the sea quarks and the unfavored
fragmentation processes. HERMES kinematics with a proton target,
COMPASS kinematics with a proton, deuteron, and neutron target (the
information on the neutron target can be effectively extracted from
the $^3$He target), and JLab kinematics (both 6~GeV and 12~GeV) with
a proton and neutron target are considered in our paper.
} 
%
\titlerunning{left-right asymmetry for pion and kaon production in the SIDIS process}
\authorrunning{Bo Sun {\it et. al.}}
\maketitle
\section{Introduction}
Single spin asymmetry (SSA) provides us with a powerful instrument
to investigate the internal structure of the nucleon and its history
can date back to the 1970s. In the early 1990s, the E704
Collaboration reported the observation of a large left-right
asymmetry in $p^\uparrow p\rightarrow \pi X$ process~\cite{E704}.
This demonstrated that the transverse spin effect is significant
even in high energies. In order to explain the unexpected
phenomenology, Sivers~\cite{Sivers} first suggested a possible
mechanism, the so-called ``Sivers effect'' today. But it was
immediately criticized by Collins~\cite{Collins1993,Collins1994},
who proposed another mechanism known as the ``Collins effect'' now.
Later, in Ref.~\cite{Anselmino1995}, by considering the soft initial
state interactions, it was argued that the Sivers effect might be
allowed. It was not until in 2002 that people began to realize that
the final state interaction plays a crucial role in leading an SSA
in the SIDIS process~\cite{Brodsky2002}. Then after considering the
gauge links, it was found that the Sivers effect, or even the Sivers
distribution could exist~\cite{Collins_etal}, and the Sivers
distribution may have different signs in the SIDIS and the Drell-Yan
processes. Despite of the early theoretical debates, some
phenomenological analysis~\cite{Anselmino2005_1,Anselmino2006}
attempted to explain the E704 data, and it was shown that the Sivers
effect is important and other effects might be suppressed. But
recently, an updated work~\cite{Anselmino2008} reported that the
Collins effect is not strongly suppressed any more after a
correction of a sign mistake. In all of these phenomenological
works, TMD factorization were assumed, but we should be aware that
the TMD factorization has not yet been proved for the $pp\rightarrow
\pi X$ process.

Contrast to the complexity of the hadron-hadron collision process,
where both the initial and final states are hadrons, the
semi-inclusive deep inelastic scattering (SIDIS) process provides a
cleaner and simpler playground for exploring the nucleon structure.
The azimuthal angle dependence of the cross section for this process
has been systematically studied in Ref.~\cite{Kotzinian1995}, where
different structure functions were defined according to different
azimuthal angle dependences. By multiplying different orthogonal
weighting functions, we can isolate different structure functions
from each other, and then extract the distribution or the
fragmentation functions from relevant terms. For example, the
Collins or the Sivers effect has a $\sin(\phi^\ell_h+\phi^\ell_S)$
or $\sin(\phi^\ell_h-\phi^\ell_S)$ \footnote{$\phi^\ell_h,
\phi^\ell_S$ are also written as $\phi_h, \phi_S$ in some other
literatures, but in this paper, we will write the explicit form with
a superscript $\ell$ to address the lepton angle dependence.}
modulation, respectively. Under the guidance, the
HERMES~\cite{hermes} and COMPASS~\cite{compass} Collaborations
studied the $\sin(\phi^\ell_h+\phi^\ell_S)$ and
$\sin(\phi^\ell_h-\phi^\ell_S)$ asymmetries, and have confirmed the
existence of the non-zero asymmetries. In the near future, JLab also
plans a high precision measurement through the SIDIS process with a
beam energy upgrading to 12~GeV. More such weighting SSAs will be
studied in these experiments, and we hope these new observations
will bring us more knowledge about the nucleon structure.

If we turn back to the E704 observation, we find that the experiment
just studied a simple un-weighted left-right asymmetry. The main
reason is that in the inclusive hadron production, only one hadron
is detected so that only one azimuthal angle can be defined. But in
the SIDIS process, both the outgoing lepton and the produced hadron
are measured, so we need to study a more complicated azimuthal
dependence involving two azimuthal angles. However, we could still
study the left-right asymmetry in the SIDIS process as the E704
experiment did. We suggest applying this method as an optional
choice in analyzing the data, for it is a simple and basic quantity.

In our previous paper~\cite{shejun}, we have studied this left-right
asymmetry for the SIDIS process in the pion production. It was
demonstrated that Sivers effect plays the most important role in
producing a left-right asymmetry for a SIDIS process. In this paper,
we will update the calculation with the new parameterizations of the
DFs and FFs, and extend the calculation to the $K^\pm$ production.
As we know, the contribution from sea quarks, especially the
$s\bar{s}$ quarks, might not be ignored in the kaon production.
Also, we will extend our calculation to more kinematics and targets
for our prediction. HERMES kinematics with a proton target, COMPASS
kinematics with a proton, deuteron, and neutron target (extracted
from the $^3$He target), and JLab kinematics (both 6~GeV and 12~GeV)
with a proton and neutron target are all considered in our paper.
The main purpose of this paper is to reproduce a left-right
asymmetry under the current theoretical framework, rather than to
study the Sivers effect extensively. In our calculation, we will use
the TMD distribution and fragmentation functions, as they may
provide us a more vivid 3-dimensional picture of a nucleon. The
proof of the TMD factorization for a SIDIS process can be found in
Ref.~\cite{factorization}. We will present our calculation up to a
leading order approximation.

\section{Definition of the asymmetry and a theoretical description}
Obviously, the asymmetry is a function of space coordinates, so it
depends on the choice of the coordinate system. For the experiments,
the most convenient choice is to choose the direction of the beam
and the target polarization as the spin plane. We call it as the
$\ell p$ frame, i.e. the laboratory frame, in which we can identify
left or right according to the spin plane. We might perform the
measurement in a space region left to the spin plane, then in its
mirror region on the right, and by their differences, we can define
a left-right asymmetry which is what the E704 experiment did. We
have noticed that changing the detected region from left ro right is
equivalence to reversing the target polarization. We can express the
asymmetry as
\begin{eqnarray}
A=-\frac{1}{S_T}\frac{N(\psi_S)-N(\psi_S+\pi)}
{N(\psi_S)+N(\psi_S+\pi)}=-\frac{1}{S_T}\frac{d\sigma^\uparrow-d\sigma^\downarrow}
{d\sigma^\uparrow+d\sigma^\downarrow}, \label{asy}
\end{eqnarray}
where $S_T$ is the transverse polarization of the target, and
$\psi_S$ is the azimuthal angle of the spin vector. The minus sign
in front of the expression is due to the fact that the detection
occurred to the right of the beam, the same as that in the E704
analysis, where no weighting function is multiplied.

Before our calculation, we first give an explanation to our
kinematics. For a theoretical description, the $\ell p$ frame is not
always convenient, since we usually regard the SIDIS process as a
virtual Compton scattering. So it is convenient to choose the
$\gamma^*p$ frame, in which the $z$ axis is defined along the
direction of the exchanged virtual photon, and the spin plane is
defined by the virtual photon and the spin vector. Thus we have two
reference frames, the $\ell p$ and the $\gamma^*p$ frames. In the
$\ell p$ frame, we can define the transverse spin vector $S_T$, the
azimuthal angle for the spin vector and the produced hadron as
$\psi_S$ and $\psi_h$, respectively. All the kinematics can be
manipulated easily in this frame, but for the theoretical
description, the $\gamma^*p$ frame might be more convenient. In the
$\gamma^*p$ frame, we could define $\phi^\ell$, $\phi_S$ and
$\phi_h$ as the azimuthal angles for the lepton plane, the spin
plain and the produced hadron plain, with respect to the horizontal
plain in the laboratory. Then we will define
$\phi_h^\ell=\phi_h-\phi^\ell,~~\phi_S^\ell=\phi_S-\phi^\ell$, and
these two angles are consistent with the Trento
convention~\cite{Trento}\footnote{These two azimuthal angles are
denoted as $\phi_h$ and $\phi_S$ in the Ref.~\cite{Trento}, but the
notations are for other uses in our paper. }.

The explicit expression for a SIDIS process can be found in
Ref.~\cite{Kotzinian1995,Bacchetta2007}, where all the coordinate
variables are defined in the $\gamma^*p$ frame. The connection
between the two frames is via a rotation by a angle $\theta$, due to
which, a transverse spin vector in the laboratory frame has a
longitudinal projection along the virtual
photon~\cite{Oganessyan2002,Diehl2005}. Generally, $\theta$ is very
small and we will make further discussion later. By taking into
account this, the cross section can be written as~\cite{Diehl2005}:
\begin{eqnarray}
&&\frac{d\sigma}{dx dy d\phi^\ell_S dz d\phi^\ell_h dP_{h\perp}^2}\nonumber\\
&&~~=\frac{\alpha^2}{2sx(1-\epsilon)}\frac{\cos\theta}{1-\sin^2\theta\sin^2\phi_S^\ell}\times
\bigg\{\mathcal{F}[f_1D_1]\nonumber\\
&&~~~~-\frac{S_T\cos\theta}{\sqrt{1-\sin^2\theta\sin^2\phi_S^\ell}}
\sin(\phi_h^\ell-\phi_S^\ell)
\mathcal{F}\bigg[\frac{\hat{\textit{\textbf{h}}}
\cdot\textit{\textbf{p}}_\perp}{M_p}f_{1T}^\perp D_1\bigg]\nonumber\\
&&~~~~-\frac{S_T\cos\theta}{\sqrt{1-\sin^2\theta\sin^2\phi_S^\ell}}
\sin(\phi_h^\ell+\phi_S^\ell)
\mathcal{F}\bigg[\frac{\hat{\textit{\textbf{h}}}
\cdot\textit{\textbf{k}}_\perp}{M_h}h_1H_1^\perp\bigg]\nonumber\\
&&~~~~+\textrm{other terms}\bigg\}\nonumber\\
&&~~\equiv
d\sigma_\textrm{UU}+d\sigma_\textrm{Siv}+d\sigma_\textrm{Col}+\ldots,
\end{eqnarray}
where we use a compact notation:
\begin{eqnarray}
\mathcal{F}[\omega fD]=\sum_a e_a^2\int d^2\textit{\textbf{p}}_\perp
d^2\textit{\textbf{k}}_\perp
\delta^2(\textit{\textbf{p}}_\perp-\textit{\textbf{k}}_\perp-\textit{\textbf{P}}_{h\perp}/z)
\nonumber\\
\omega(\textit{\textbf{p}}_\perp,\textit{\textbf{k}}_\perp)
f^a(x,p^2_\perp)D^a(z,z^2k^2_\perp),~~~
\end{eqnarray}
and
\begin{eqnarray}
\epsilon=\frac{1-y-\frac{1}{4}y^2\gamma^2}{1-y+\frac{1}{2}y^2+\frac{1}{4}y^2\gamma^2},~~~~~
\hat{\textit{\textbf{h}}}\equiv\textit{\textbf{P}}_{h\perp}/|\textit{\textbf{P}}_{h\perp}|.
\end{eqnarray}

First, we make a first approximation that $\theta$ is small (We will
give a detailed discussion later), thus the $\ell p$ and $\gamma^*p$
frames are of no difference. Now under this approximation, we can
use the kinematics defined in the $\gamma^*p$ frame instead to
analyze the asymmetry defined in Eq.~\ref{asy},
\begin{eqnarray}
A_{UT}(x,y,z,P_{h\perp})\approx-\frac{1}{S_T}\frac{\int d\phi^\ell_S
d\phi^\ell_h
~(d\sigma_\textrm{Siv}+d\sigma_\textrm{Col}+\ldots)}{\int
d\phi^\ell_S d\phi^\ell_h ~d\sigma_\textrm{UU}}.
\end{eqnarray}
Next, we change the integral measure from $d\phi^\ell_S
d\phi^\ell_h$ to $d\phi^\ell d\phi_h$ (The jacobian $|J|=1$), and
perform the integral over $\phi^\ell$. We notice that
$\sin\phi_S^\ell$ explicitly depends on $\phi^\ell$ and all the
convolution integrals $\mathcal{F}$ are independent of $\phi^\ell$.
For the azimuthal angle dependence, we find that all the factors are
oscillation functions of $\phi^\ell$ except $\sin(\phi_h-\phi_S)$,
which is $\phi^\ell$-independent. So after integrating over
$\phi^\ell$, we find that Sivers effect is $o(1)$, but other terms
such as the Collins effect are $o(\sin^2\theta)$ (See detailed
discussion in Ref.~\cite{shejun}). So Sivers effect is dominant and
other effects are suppressed in our analysis. We could also
understand this result as the following. If we ignore the angle
$\theta$, we could set the $\gamma^*p$ frame equal to the $\ell p$
frame in the laboratory. In the E704 method, only one azimuthal
angle was involved in fact, i.e. $\phi_h$ in our notation (not
$\phi_h^\ell$), and the lepton angle $\phi^\ell$ in not included in
the analysis. For a SIDIS process, only the Sivers effect is
independent of the lepton plane, so it is not strange that the
Sivers effect plays the most important role in leading an left-right
asymmetry.

We could make an estimation on the effect resulted from the angle
$\theta$. This angle can be calculated from the kinematical
variables~\cite{Diehl2005}
\begin{eqnarray}
\sin\theta=\gamma\sqrt{\frac{1-y-\frac{1}{4}y^2\gamma^2}{1+\gamma^2}},~~~~~
\gamma=2xM_p/Q .
\end{eqnarray}
We replace each variable by its average value to estimate the mean
value of $\sin\theta$ for the HERMES and JLab experiments.
\begin{table}
\caption{An estimation on $\sin\theta$} \label{theta}
\begin{center}
\begin{tabular}{c|c|c|c}
\hline\hline
&HERMES&\multicolumn{2}{|c}{JLab}\\
\cline{3-4}
&&6~GeV&12~GeV\\
\hline
$\langle x\rangle$&0.09&0.23&0.23\\
$\langle y\rangle$&0.54&0.6&0.57\\
$\langle Q^2\rangle$&2.41~GeV$^2$&1.8~GeV$^2$&2.5~GeV$^2$\\
$\langle \sin\theta\rangle$&0.073&0.19&0.17\\
\hline\hline
\end{tabular}
\end{center}
\end{table}

The estimated result is shown in Table~\ref{theta}, and we find that
for most instance, the direction of the virtual photon is very close
to the direction of the incident beam. Therefore, for convenience,
the Collins effect which is not known so clearly yet is not
considered in our analysis.

\section{Parametrization for distribution and fragmentation
functions}

As a preparation for our calculation, we will present the
parametrization for the distribution and fragmentation functions we
will use in this section.

For the Sivers functions, there are already some model
calculations~\cite{Sivers_func_model}, but we would use a
phenomenological parameterization for the Sivers functions. We
should be cautious that a universal transverse momentum dependent
Sivers distribution for different processes does not
exist~\cite{Bomhof08}. Fortunately, we will calculate for the SIDIS
process, and the parametrization of the Sivers function is also from
the SIDIS data. The Sivers effect has already been studied by HERMES
and COMPASS Collaborations, and extractions on the Sivers functions
for the $u$ and $d$ quarks were already
obtained~\cite{Anselmino2005_1,Anselmino2005_2,Efremov2005,Collins2006}.
But all these results were under low statistics and assumed the
existence of a symmetric and negligibly small Sivers sea. Recently,
the HERMES Collaboration has provided much higher statistic data on
the $A_{UT}^{\sin(\phi_{h}-\phi_{S})}$ azimuthal
asymmetry~\cite{hermesnew}. Besides the charged pion production,
neutral pion and charged kaon azimuthal asymmetries were also
analyzed. Also, the COMPASS Collaboration separated the charged pion
and charged kaon asymmetries from the charged hadron production
measurement~\cite{comapss,compassnew}. These SIDIS experimental data
on the Sivers asymmetries for the pion and kaon production give us
an opportunity to study the sea-quark Sivers functions for the $\bar
u$, $\bar d$, $s$, and $\bar s$ quarks. With these data, in
Ref.~\cite{Anselmino2009}, the extraction of these functions was
improved and the first estimates of the sea-quark Sivers functions
were presented. The Sivers function is parameterized in the form
\begin{eqnarray}
&&f^{\perp q}_{1T}(x,p^2_\perp)=-\frac{M_p}{p_\perp}\mathcal{N}_q(x)
f_q(x)g(p^2_\perp)h(p^2_\perp),\\
&&\mathcal{N}_q(x)=N_q x^\alpha_q (1-x)^\beta_q
\frac{(\alpha_q+\beta_q)^{(\alpha_q+\beta_q)}}{\alpha_q^{\alpha_q} \beta_q^{\beta_q}},\\
&&g(p^2_\perp)=\frac{e^{-p^2_\perp/\langle p^2_\perp \rangle}}{\pi
\langle p^2_\perp
\rangle},~~~h(p^2_\perp)=\sqrt{2e}\frac{p_\perp}{M'}e^{-p^2_\perp/\langle
M'^2 \rangle}.
\end{eqnarray}

\begin{figure}
\begin{center}
\resizebox{0.4\textwidth}{!}{%
  \includegraphics{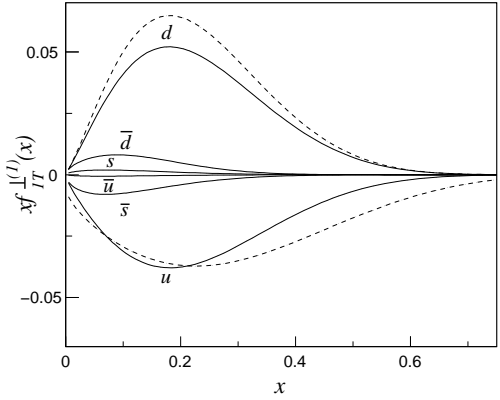}
} \caption{Sivers functions for different quark flavors. Solid
curves are the new results according to Ref.~\cite{Anselmino2009},
and dashed curves are that according to Ref.~\cite{Anselmino2005_2}}
\label{xf1t}
\end{center}
\end{figure}

All the parameters can be found in Ref.~\cite{Anselmino2009}. In the
above parametrization, $f(x)$ is the unpolarized parton distribution
functions, and we adopt the CTEQ6L parametrization~\cite{CTEQ6} as
an input. We plot the Sivers functions for different quark flavors
in Fig.~\ref{xf1t}

\begin{figure}
\center \resizebox{0.4\textwidth}{!}{\includegraphics{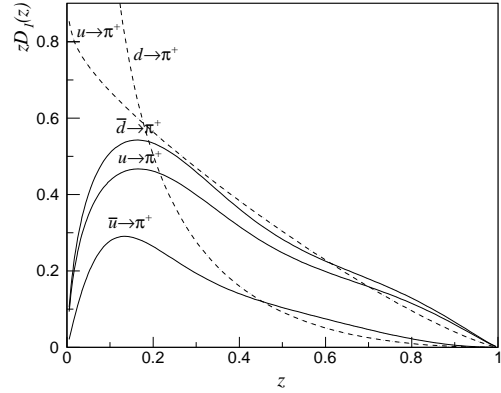} }
\caption{Fragmentation functions for $\pi^+$. Solid curves are the
results according to Ref.~\cite{Florian2007}, and dashed curves are
the results used in Ref.~\cite{shejun}.} \label{pifrag}
\end{figure}

\begin{figure}
\center \resizebox{0.4\textwidth}{!}{\includegraphics{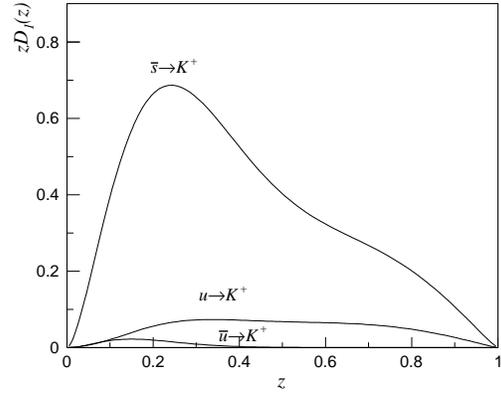} }
\caption{Fragmentation functions for $K^+$ according to
Ref.~\cite{Florian2007}} \label{kfrag}
\end{figure}

For the fragmentation functions, all the former analysis of the
fragmentation functions were based exclusively on the
single-inclusive $e^+ e^-$ annihilation (SIA) data and have been
chosen the most simple functional form  $N_i z^{\alpha_i}
(1-z)^{\beta_i}$ to parametrize the $D_i^H$ . But in these
experiments, information on the quark and anti-quark fragmentation
is always combined, for it always refers to the charge sum for
certain hadron species, e.g. $\pi^++\pi^-$. In order to distinguish
``valence'' from ``sea'' fragmentation, some assumptions were
proposed, e.g., in Ref.~\cite{Kretzer2000},
$D_{\bar{u}}^{\pi^+}/D_{u}^{\pi^+}=(1-z)$ was assumed. In the last
few years several one-particle inclusive measurements coming from
both the proton-proton collisions and the deep-inelastic
lepton-nucleon scattering gave an opportunity to weigh each quark
contribution in the hadronization process. In
Ref.~\cite{Florian2007}, a global analysis was taken for the first
time to analyze the individual fragmentation functions for all
flavors as well as gluons. A more flexible input is used
\begin{equation}
\label{eq:ff-input} D_i^H(z,\mu_0) = \frac{N_i
z^{\alpha_i}(1-z)^{\beta_i} [1+\gamma_i (1-z)^{\delta_i}] }
{\textrm{B}[2+\alpha_i,\beta_i+1]+\gamma_i
\textrm{B}[2+\alpha_i,\beta_i+\delta_i+1]},
\end{equation}
where $\textrm{B}[a,b]$ is the Beta-function and $N_i$ is normalized
to represent the contribution of $D_i^H$ to the sum rule. For the
fragmentation to $\pi^+$, the isospin symmetry for the sea
fragmentation functions is imposed, i.e.,
\begin{equation}
\label{eq:iso} D_{\bar{u}}^{\pi^+}=D_{d}^{\pi^+}.
\end{equation}
But slightly different normalization in the $q+\bar{q}$ sum is
allowed:
\begin{equation}
\label{eq:val_break} D_{d+\bar{d}}^{\pi^+}= N D_{u+\bar{u}}^{\pi^+}.
\end{equation}
For the strange quarks it is assumed that
\begin{equation}
\label{eq:sea_break} D_s^{\pi^+}=D_{\bar{s}}^{\pi^+} =N^{\prime}
D_{\bar{u}}^{\pi^+}.
\end{equation}
For the charged kaons, it is assumed that the unfavored
fragmentation functions are the same,
\begin{equation}
\label{eq:sea_ka}
D_{\bar{u}}^{K^+}=D_{s}^{K^+}=D_d^{K^+}=D_{\bar{d}}^{K^+}.
\end{equation}
>From these relations, the unfavored fragmentation functions can be
distinguished form the favored ones. Detailed parametrization for
the integrated fragmentation function $D(z)$ can be found in
Ref.~\cite{Florian2007}. We present the numerical results for the
charged pion and kaon fragmentation functions in Fig.~\ref{pifrag}
and Fig.~\ref{kfrag}. We notice that in this parametrization, the
$\bar{s}\rightarrow K^+$ process is the most favored one, which
means that sea quarks, especially $\bar{s}$ quark, might contribute
significantly, although sea quark distributions are small compared
with the valence quarks. So measurements on the kaon production may
help us to know more about the $s(\bar{s})$ distribution. In our
calculation, we need the TMD fragmentation function, and we adopt a
Gaussian assumption
\begin{eqnarray}
D_1(z,z^2k^2_\perp)=D_1(z)\frac{\exp(-z^2{k}_\perp^2/R^2)}{\pi R^2},
\end{eqnarray}
with $R^2=0.2~\mathrm{GeV}^2$ suggested in
Ref.~\cite{Anselmino2005_3}.


\section{Numerical calculations}

The kinematical cuts used in the calculation are shown in
Table~\ref{kin}. For the HERMES experiment, only the proton target
is calculated. For the Compass experiment, the proton, neutron and
deuteron targets are all considered, while for the JLab experiment,
the proton and neutron targets are assumed. In fact there is no free
neutron target, and in experiments the polarized $^3$He is used. The
effective asymmetry on a free neutron can be extracted from a $^3$He
target asymmetry. Detailed discussions can be found in some
theoretical works~\cite{He3} and a JLab's proposal~\cite{JLab}. We
will investigate the $x$ and $z$ dependence\footnote{The E704
experiment only showed the dependence on $x_F$, i.e. approximate $z$
here.} of the asymmetries.
\begin{table*}
\caption{Kinematics} \label{kin}
\begin{center}
\begin{tabular}{c|c|c|c|c|c|c}
\hline\hline
&HERMES&COMPASS&\multicolumn{2}{|c|}{JLab1}&\multicolumn{2}{|c}{JLab2}\\
\cline{4-7}
&&&proton&neutron&proton&neutron\\
\hline
$p_\textrm{beam}$/GeV&27.6&160&6&6&12&12\\
$Q^2$/GeV$^2$&$>1$&$>1$&$>1$&$1.3\sim3.1$&$>1$&$>1$\\
$W^2$/GeV$^2$&$>10$&$>25$&$>4$&$5.4\sim9.3$&$>4$&$>2.3$\\
$x$&$0.023\sim0.4$&$$&$0.1\sim0.6$&$0.13\sim0.4$&$0.05\sim0.7$&$0.05\sim0.55$\\
$y$&$0.1\sim0.85$&$0.1\sim0.9$&$0.4\sim0.85$&$0.68\sim0.86$&$0.2\sim0.85$&$0.34\sim0.9$\\
$z$&$0.2\sim0.7$&$0.2\sim1$&$0.4\sim0.7$&$0.46\sim0.59$&$0.4\sim0.7$&$0.3\sim0.7$\\
\hline\hline
\end{tabular}
\end{center}
\end{table*}

\begin{figure}
\begin{center}
\resizebox{0.4\textwidth}{!}{\includegraphics{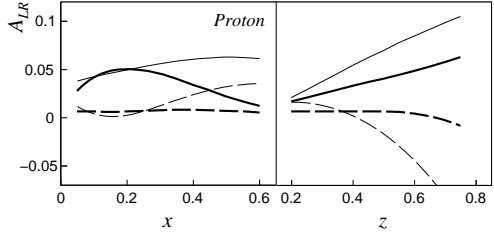} }
\end{center}
\caption{The $x$ and $z$-dependence of the left-right asymmetry for
$\pi^{\pm}$ production on HERMES kinematics. Solid lines for $\pi^+$
and dashed lines for $\pi^-$. Thick curves are our results and thin
curves are results from Ref.~\cite{shejun}} \label{hermespi}
\end{figure}

\begin{figure}
\begin{center}
\resizebox{0.4\textwidth}{!}{\includegraphics{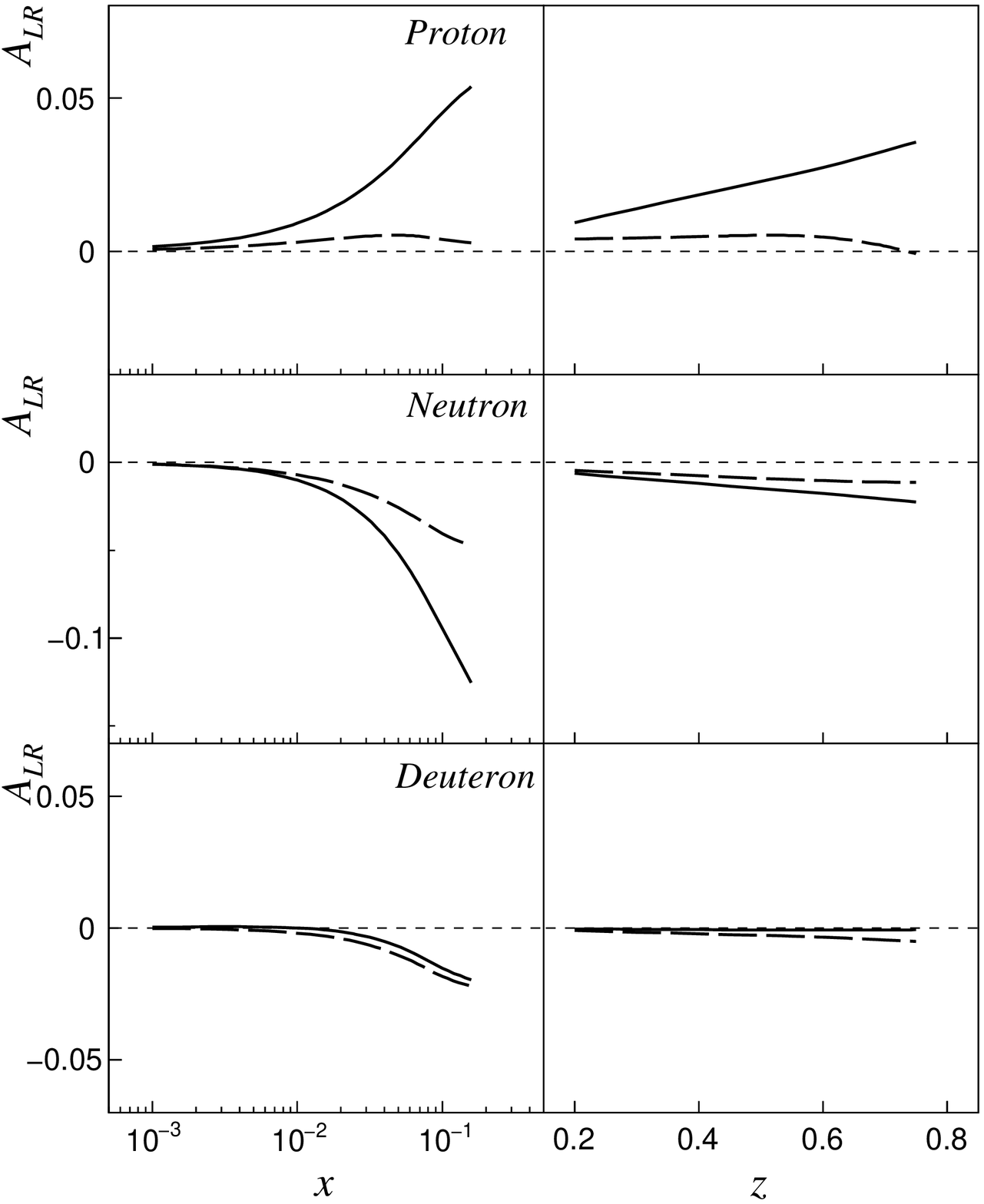} }
\end{center}
\caption{Similar as Fig.~\ref{hermespi}, but at COMPASS kinematics.}
\label{compasspi}
\end{figure}

\begin{figure}
\center \resizebox{0.4\textwidth}{!}{\includegraphics{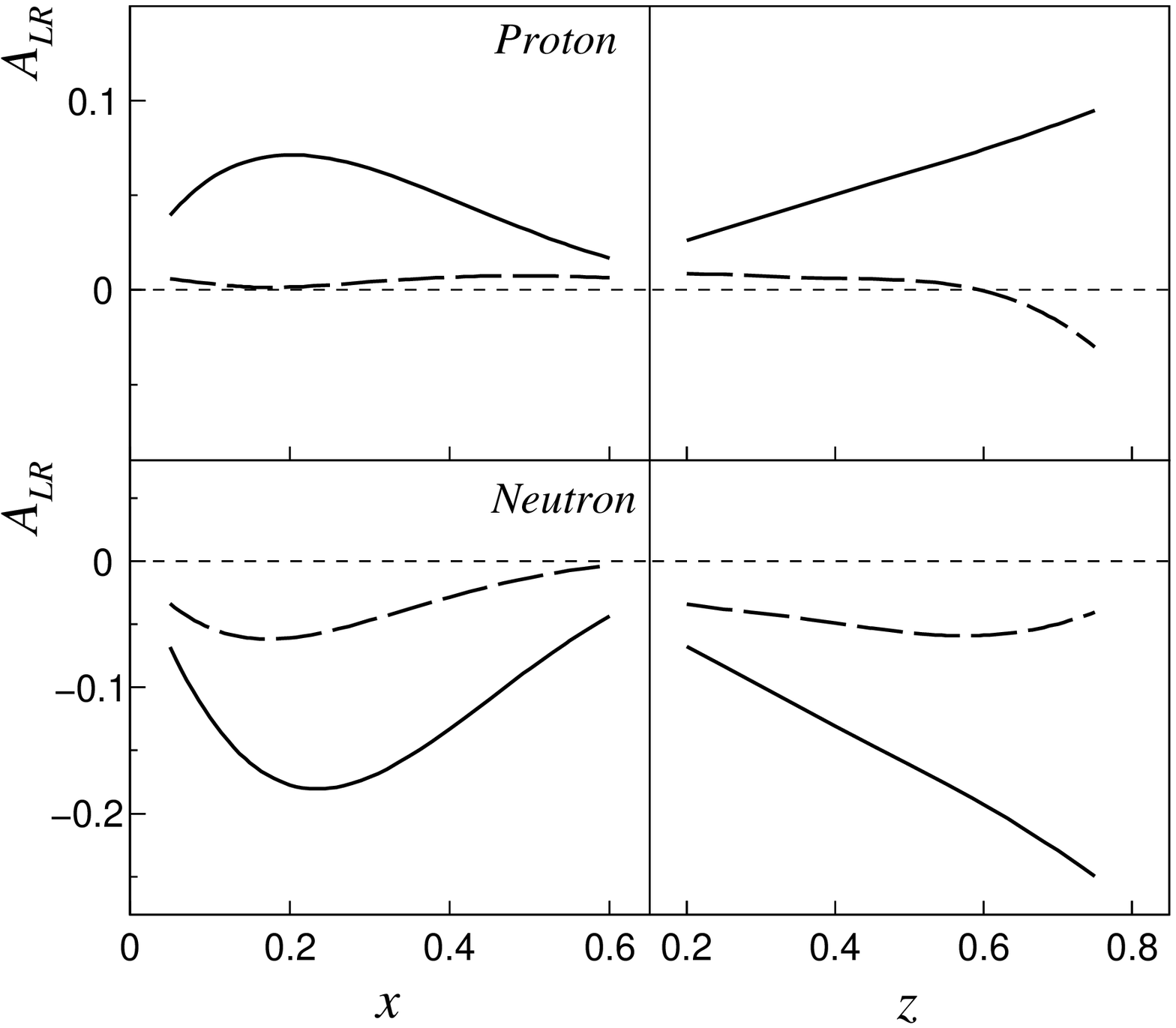} }
\caption{Similar as Fig.~\ref{hermespi}, but at JLab kinematics with
a beam energy of 6 GeV.} \label{jlab6pi}
\end{figure}

\begin{figure}
\begin{center}
\resizebox{0.4\textwidth}{!}{\includegraphics{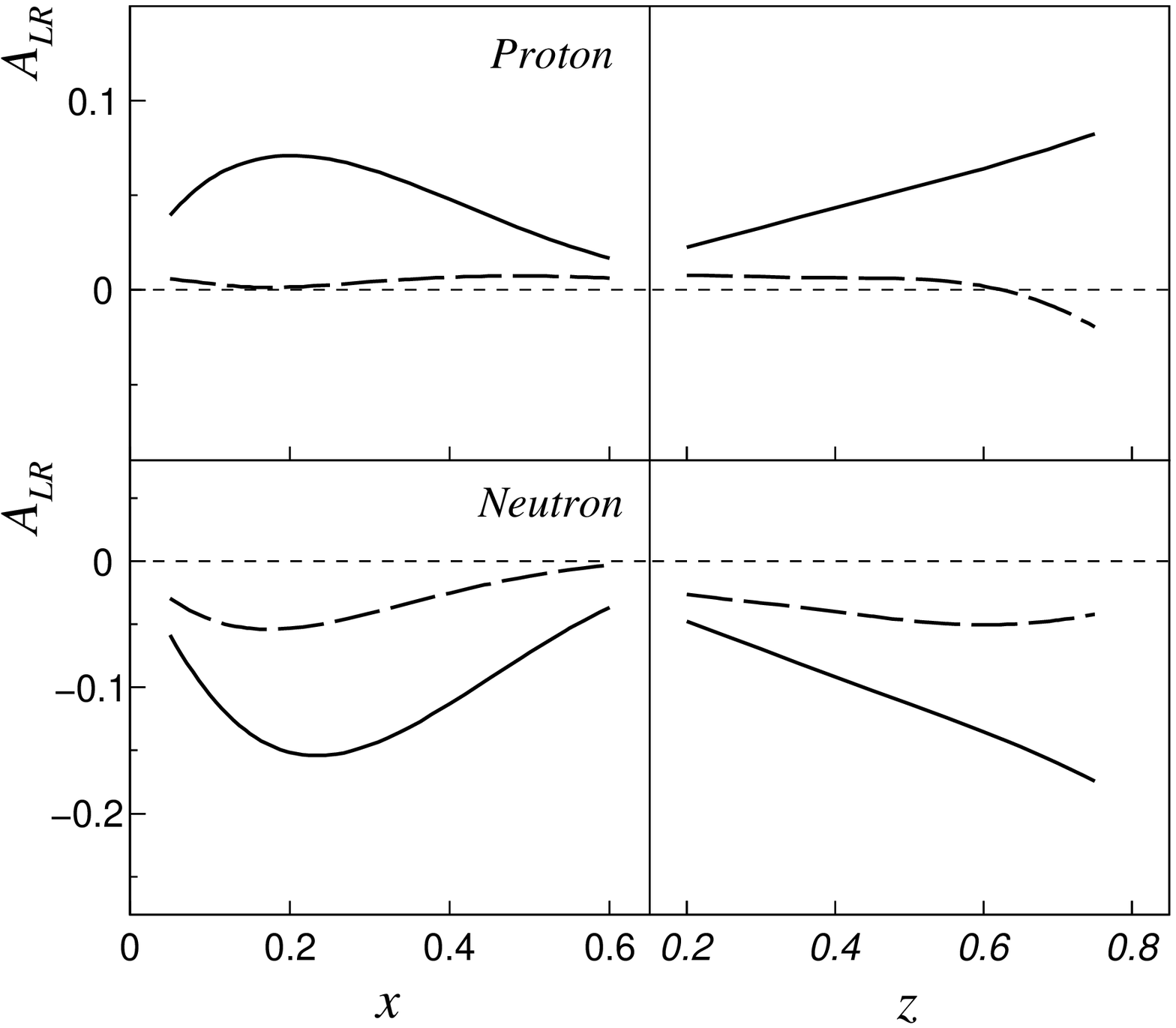} }
\end{center}
\caption{Similar as Fig.~\ref{hermespi}, but at JLab kinematics with
a beam energy of 12 GeV.} \label{jlab12pi}
\end{figure}

\begin{figure}
\begin{center}
\resizebox{0.4\textwidth}{!}{\includegraphics{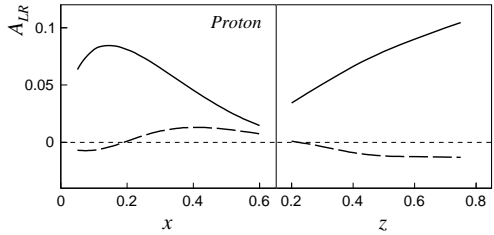} }
\end{center}
\caption{The $x$ and $z$-dependence of the left-right asymmetry for
$K^{\pm}$ production on HERMES kinematics. Solid lines for $K^+$ and
dashed lines for $K^-$.} \label{hermesk}
\end{figure}

\begin{figure}
\begin{center}
\resizebox{0.4\textwidth}{!}{\includegraphics{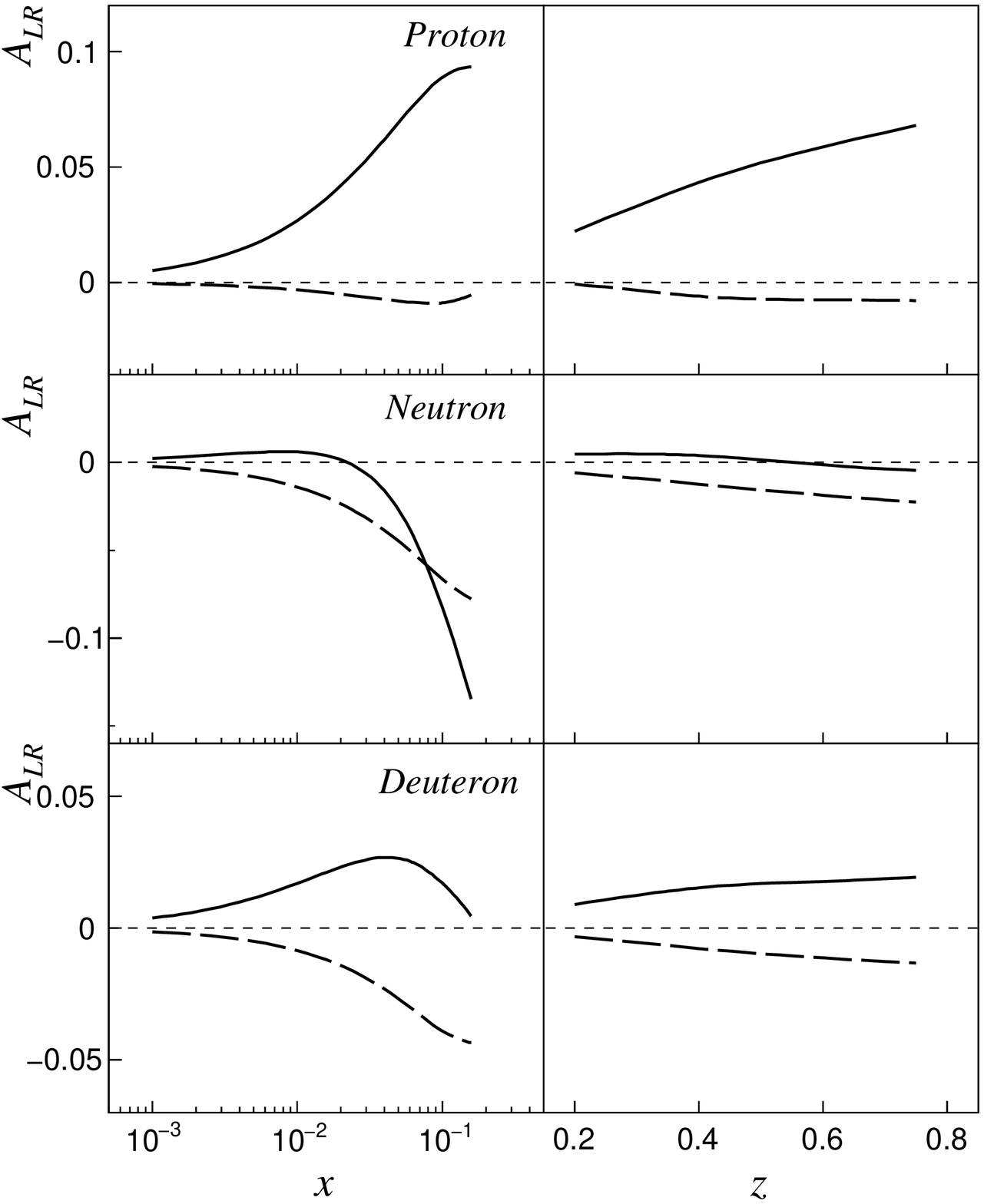} }
\end{center}
\caption{Similar as Fig.~\ref{hermesk}, but at COMPASS kinematics.}
\label{compassk}
\end{figure}

\begin{figure}
\center \resizebox{0.4\textwidth}{!}{\includegraphics{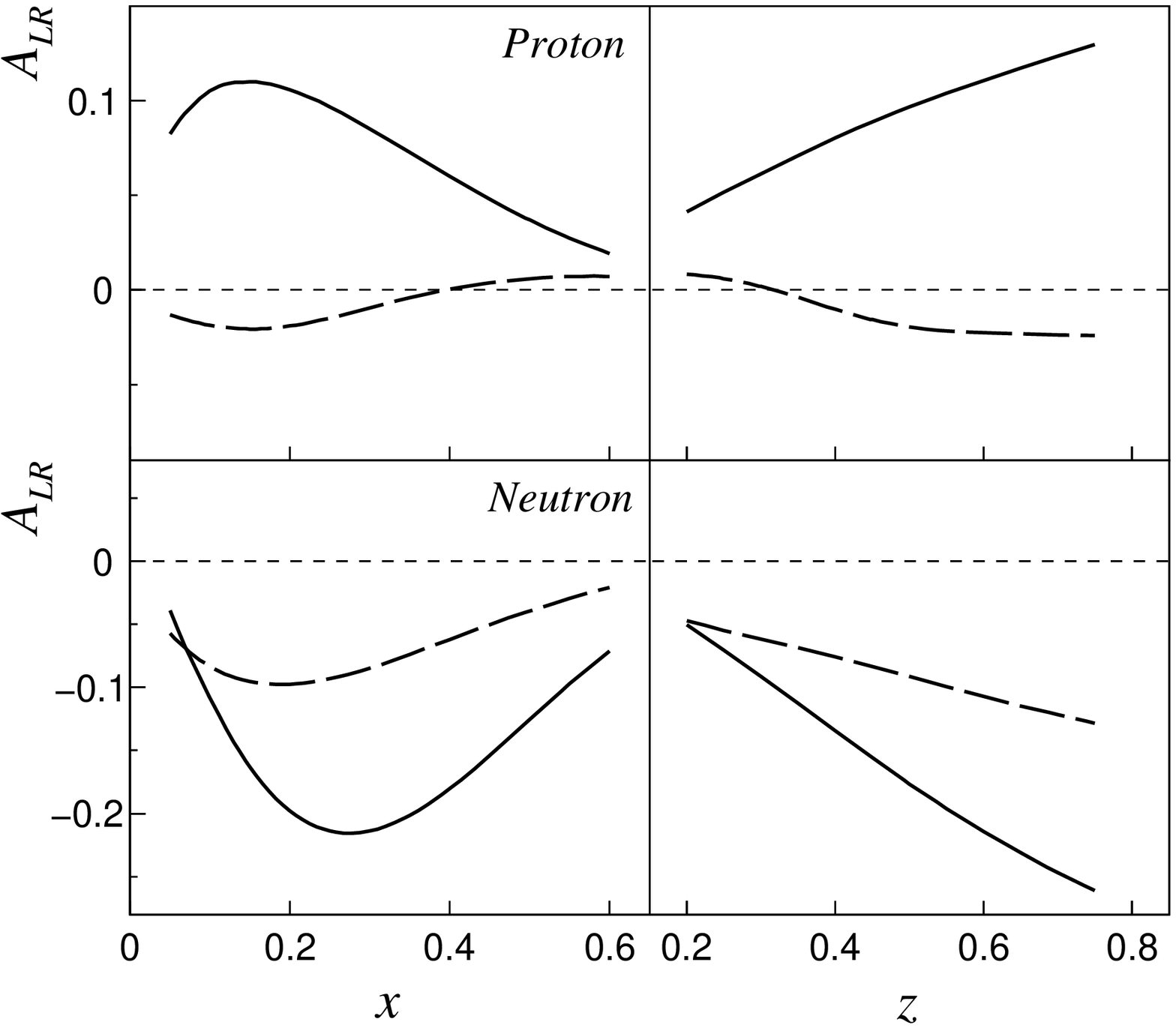} }
\caption{Similar as Fig.~\ref{hermesk}, but at JLab kinematics with
a beam energy of 6 GeV.} \label{jlab6k}
\end{figure}

\begin{figure}
\center \resizebox{0.4\textwidth}{!}{\includegraphics{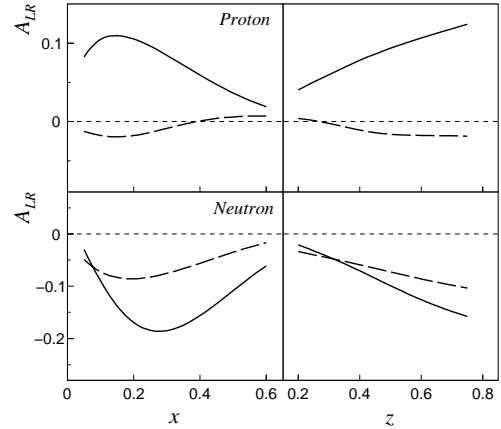} }
\caption{Similar as Fig.~\ref{hermesk}, but at JLab kinematics with
a beam energy of 12 GeV.} \label{jlab12k}
\end{figure}

First, we present our results on $\pi^{\pm}$ production at different
kinematics. Fig.~\ref{hermespi} - Fig.~\ref{jlab12pi} show the
results of the left-right asymmetry. In Fig.~\ref{hermespi}, we make
a comparison with the results already obtained in
Ref.~\cite{shejun}, and we find that the two results are a little
different. For the $x$-dependence of the asymmetry, our new results
are suppressed when $x$ increases. We can find the reason from the
parametrization for Sivers functions as Fig.~\ref{xf1t} shows. The
new parametrization shows that at large $x$ region ($x>0.3$), the
$u$ quark distribution falls down much faster than a previous
parametrization, while the $d$ quark distribution does not change so
much. Notice that the $u$ quark distribution gives a positive
contribution, then we could understand our new results. For the
$z$-dependence, the difference results not only from the
distributions, but also from different parametrizations of the
fragmentation functions. So we can say that our results are
sensitive to the parametrization.

Next we will extend our calculation to the $K^{\pm}$ production, and
the results are shown in Fig.~\ref{hermesk} - Fig.~\ref{jlab12k}.
The $K^+$ production is quite similar to that of $\pi^+$, but for
the $K^-$ production, we should be cautious. $K^-$ is made up of a
$\bar{u}$ and an $s$ quark. So the sea quark contribution (from
$\bar{u}$ and $s$) might be enhanced due to the favored
fragmentation process, while the valence quark contribution would be
suppressed. Thus, the $K^-$ production is a good way to study the
sea quark distributions and the unfavored fragmentation processes.
Notice that the Sivers distributions of the $\bar{u}$ and $s$ quarks
given in Ref.~\cite{Anselmino2009} are so small with a large
uncertainty that even their signs are not determined within the
error, so our prediction must cover a large uncertainty area.
Nevertheless, we hope that our results will be helpful to the future
experiments and we expect that further experiments with a high
precision will clarify the detail.

\section{Conclusion}
Following the E704 analysis, we reanalyzed the left-right asymmetry
in the SIDIS process with the new Sivers functions and fragmentation
functions. In this paper, we considered all the flavor
contributions, including the sea quarks. We extended our analysis to
the $K^\pm$ production process, and meanwhile to various kinematics
with different targets. We found that our results are sensitive to
the parametrization form of the distribution and fragmentation
functions, so we consider it necessary to perform higher precision
measurements to constrain the parametrization.

Our prescription originated from the E704 experiment is an optional
and simple way analyzing the data. In this prescription, no
weighting functions are multiplied, although it might not give any
more information. We suggest that relevant experiment collaborations
could present their data in this new way as an optional choice for
further theoretical studies.

\section*{Acknowledgement}
This work is partially supported by National Natural Science
Foundation of China (No. 10375002, No. 10675004, No.~10721063 and
No.~10975003), by the Key Grant Project of Chinese Ministry of
Education (No.~305001), by the Research Fund for the Doctoral
Program of Higher Education (China).

\end{document}